\documentclass{PoS}
\usepackage{amsmath}
\usepackage{stackrel}
\usepackage{slashed}
\usepackage{mathrsfs}
\usepackage{array}
\newcolumntype{P}[1]{>{\centering\arraybackslash}p{#1}}

\title{Computing the nucleon Dirac radius directly at $\mathbf{Q^2=0}$}

\ShortTitle{Computing the nucleon Dirac radius directly at $Q^2=0$}%Nucleon Dirac radius directly at $Q^2=0$}

\author{\speaker{Nesreen Hasan}$^{ab}$, Michael Engelhardt$^d$, Jeremy Green$^{g}$\thanks{Current affiliation: NIC, DESY, Zeuthen, Germany}, Stefan Krieg$^{ab}$, Stefan~Meinel$^{eh}$, John Negele$^c$, Andrew Pochinsky$^{c}$ and Sergey Syritsyn$^{fh}$ \\
\llap{$^a$} Bergische Universit\"at Wuppertal, Wuppertal, Germany \\
\llap{$^b$} JSC, J\"ulich Supercomputing Centre, Forschungszentrum J\"ulich, J\"ulich, Germany\\
\llap{$^c$} Center for Theoretical Physics, Massachusetts Institute of Technology, Cambridge, USA\\
\llap{$^d$} Department of Physics, New Mexico State University, Las Cruces,  NM 88003, USA\\
\llap{$^e$} Department of Physics, University of Arizona, Tucson, AZ 85721, USA\\
\llap{$^f$}  Department of Physics and Astronomy, Stony Brook University, Stony Brook, NY 11794, USA \\
\llap{$^g$} PRISMA Cluster of Excellence and Institut f\"ur Kernphysik, Johannes Gutenberg-Universit\"at Mainz, Germany\\
\llap{$^h$} RIKEN BNL Research Center, Brookhaven National Laboratory, Upton, NY 11973, USA
E-mail: \email{n.hasan@fz-juelich.de},  \email{engel@nmsu.edu}, \email{ jeremy.green@desy.de}, \email{s.krieg@fz-juelich.de},  \email{smeinel@email.arizona.edu}, \email{negele@mit.edu}, \email{avp@mit.edu}, \email{ssyritsyn@quark.phy.bnl.gov}  
}
\abstract{We describe a lattice approach for directly computing momentum derivatives
 of nucleon matrix elements using the Rome method, which we apply
to obtain the isovector magnetic moment and Dirac radius. We present preliminary
 results calculated at the physical pion mass using a 2HEX-smeared
Wilson-clover action. For removing the effects of excited-state contamination, the calculations
were done at three source-sink separations and the summation method was
used.
}

\FullConference{34th annual International Symposium on Lattice Field Theory\\
		24-30 July 2016\\
		University of Southampton, UK}

\begin{document}
%-------------------
\section{Introduction}
%---------------------
The ``proton radius puzzle'' refers to the $7\sigma$ discrepancy in the experimental determinations of the charge radius of the proton between the value extracted from spectroscopy of muonic hydrogen \cite{b}, $r_E^p=0.84087(39)$ fm, and the CODATA value determined from scattering and spectroscopy with electrons \cite{c}, $r_E^p = 0.8775(51)$ fm. 
The way to determine the proton radius $r_E^p$ from scattering data is to measure the slope of the electric form factor $G_E^p(Q^2)$ as a function of the squared-momentum transfer $Q^2$ as $Q^2$ approaches zero. There is controversy surrounding finding the radius from fitting the scattering data: some claim to find a small radius from scattering data  \cite{d,e,f} and others argue that only a large radius is compatible with the data \cite{g,h,y}. 
On the lattice, the same approach is usually followed. However, the smallest nonzero $Q^2$ reached on the lattice 
with standard methods and a large volume of $(5.6 \text{\;fm})^3$ is about $0.05\;\text{GeV}^2$, whereas the smallest $Q^2$ reached in scattering experiments is below $0.005\; \text{GeV}^2$. 
 Therefore, obtaining a reliable radius from fitting to lattice form factor data may be challenging. This motivates the need for a direct calculation of the radius without fitting to form factors.
For the case of a pion, it was shown in \cite{o} that the Rome method for momentum derivatives could be used to calculate the pion charge radius with finite-volume effects that are exponentially suppressed, with asymptotic behaviour $\sim \sqrt{m_\pi L} \;e^{-m_\pi L}.$

In the following, we present a method for calculating the nucleon Dirac radius directly at $Q^2 =0$. Our approach is based on calculating the momentum derivatives of the two- and three-point functions using the Rome method \cite{a}, as explained in section \ref{der}.
 A ratio of the two- and three-point functions is then constructed and by calculating the momentum derivatives of this ratio at $Q^2=0$ we are able to extract the anomalous magnetic moment $\kappa=F_2(0)$  and Dirac radius  $r_1^2 = \frac{-6}{F_1}\;\frac{dF_1}{dQ^2}\Big|_{Q^2=0}$ (sections \ref{Sd} and  \ref{MD}). Our results for both quantities are shown in section \ref{LC}.
 %--------
\section{Momentum derivatives of the two-point and three-point functions}
\label{der}
For calculating the  momentum derivatives of the correlation functions, we need to consider quark propagators with smeared- and point-sources and sinks. This is shown in the following:\\ \\
%-------------------
\textbf{Without smearing}: 
%\label{ns}
%---------------------
Let $N$ denote a proton or neutron field. The two-point function can be written as:
\small
\begin{align}
C_2(\mathbf p, x^0,z^0)_{\alpha\beta} &= \sum_{\mathbf x} e^{-i\mathbf{p(x-z)}} \langle N_\alpha(x) \bar N_{\beta}(z)\rangle \nonumber \\
&= \sum_{\mathbf x} e^{-i\mathbf{p(x-z)}} \epsilon^{abc}\epsilon^{def} f_{\alpha\gamma\delta\epsilon} \bar f_{\beta\zeta\eta\theta}  \big\langle G_{\gamma\theta}^{af}(x,z) G_{\delta\eta}^{be}(x,z)G_{\epsilon\zeta}^{cd}(x,z) - G_{\gamma\eta}^{ac}(x,z)G_{\delta\theta}^{bf}(x,z) G_{\epsilon\zeta}^{cd}(x,z)\big \rangle\nonumber\\
&=\sum_{\mathbf x} \epsilon^{abc}\epsilon^{def} f_{\alpha\gamma\delta\epsilon} \bar f_{\beta\zeta\eta\theta} \big \langle G_{\gamma\theta}^{af}(x,z) G_{\delta\eta}^{be}(x,z)G_{\epsilon\zeta}^{cd}(x,z;\mathbf{p}) - G_{\gamma\eta}^{ac}(x,z)G_{\delta\theta}^{bf}(x,z) G_{\epsilon\zeta}^{cd}(x,z;\mathbf{p})\big \rangle,
\end{align}
\normalsize
%= (\frac{1+\gamma^0}{2})_{\alpha\delta} (C \gamma_5 \frac{1+\gamma^0}{2})_{\gamma\beta}
where $f_{\alpha\beta\gamma\delta}$ is the spin tensor determining the quantum numbers of the nucleon operator $N$, $G(x,z)$ is the quark propagator and $G(x,z;\mathbf p) = e^{-i\mathbf{p(x-z)}} G(x,z)$. As shown in \cite{a}, the first and second momentum derivatives of a quark propagator at zero momentum are given by:
\begin{align}
\frac{\partial}{\partial p^j} G(x,y;\mathbf p) \big|_{\mathbf p=0} &= -i\sum_z G(x,z) \Gamma_V^j G(z,y), \label{fmd}\\
\frac{\partial^2}{(\partial p^j)^2} G(x,y;\mathbf p) \big|_{\mathbf p=0} &= -2\sum_{z,z'} G(x,z) \Gamma_V^j G(z,z') \Gamma_V^j G(z',y) -\sum_z G(x,z) \Gamma_T^j G(z,y), \label{smd}
\end{align}
where % the sum is over the four-dimensional coordinates $z, z'$,% the `` point-split'' and ``tadpole'' operators are
\begin{equation}
\Gamma_{V/T}^j G(z,y) \equiv U_j^\dagger(z-\hat{\jmath}) \frac{1+\gamma^j}{2} G(z-\hat{\jmath},y) \mp U_j(z) \frac{1-\gamma^j}{2}G(z+\hat{\jmath},y).
\end{equation}
For connected diagrams, the three-point function, with current $O_\Gamma=\bar q \Gamma q$  and zero sink momentum $\mathbf p' =0$, can be written as:
\begin{equation}
C_3(\mathbf p,x^0,y^0,z^0)_{\alpha\beta} = \sum_{\mathbf{x,y}} e^{-i\mathbf{p(y-z)}} \langle N_\alpha (x) O_\Gamma (y) \bar N_\beta(z)\rangle  =\sum_{\mathbf y} \langle G_S(y) \Gamma G(y,z;\mathbf p)\rangle, 
\end{equation}
where $G_S(y)$ is the sequential backward propagator, which is independent of $\mathbf p$. Only the forward propagator $G(y,z;\mathbf p)$ needs to be expanded using \ref{fmd} and \ref{smd}. Hence, no additional backward propagators are needed.
The Rome method can be understood as doing a calculation with twisted boundary conditions and then taking the derivative with respect to the twist angle at zero twist \cite{o}. \\\\
%------------------------------------
\textbf{With smearing}:
%\label{ws}
%-------------------------------------
In the two-point function, we have the smeared-source smeared-sink propagator
%We can write the propagator with momentum $\mathbf{p}$ and with smeared source and smeared sink with a smearing kernel $K$ as follows:
%In the two-point function, we have smeared-source smeared-sink propagator
\small
\begin{align}
\tilde{\tilde{G}}(x,y;\mathbf p) &= e^{-i\mathbf{p(x-y)}} \sum_{x',y'} K(x,x') G(x',y') K(y',y) \nonumber \\
&=\sum_{x',y'} \underbrace{e^{-i\mathbf{p(x-x')}} K(x,x')}_{ K(x,x';\mathbf p)}  \underbrace{e^{-i\mathbf{p(x'-y')}} G(x',y')}_{G(x',y';\mathbf p)}  \underbrace{e^{-i\mathbf{p(y'-y)}} K(y',y)}_{ K(y',y;\mathbf p)} \nonumber,
%&= \sum_{x',y'} K(x,x';\mathbf p) G(x',y';\mathbf p) K(y',y;\mathbf p),
\end{align}
\normalsize
where $K$ is the smearing kernel. The momentum derivatives can then be calculated using the product rule along with \ref{fmd} and  \ref{smd}. Denoting the momentum derivative with $'$ for shorter notation, \\
\small
\noindent\begin{minipage}{.4\linewidth}
\begin{align}
 (KGK)' = K'GK + K(GK)' , \\ \nonumber
\end{align}
\end{minipage}
\begin{minipage}{.6\linewidth}
\begin{align}
(KGK)'' = K'' GK +2K'(GK)'+K(GK)''. \\ \nonumber
\end{align}
\end{minipage}
\normalsize
\small
For the smeared-source point-sink propagator, which is needed for the three-point function, we get:\\
\noindent\begin{minipage}{.4\linewidth}
\begin{align}
 (GK)' = G[-i\Gamma_V G K + K'], \\ \nonumber
\end{align}
\end{minipage}
\begin{minipage}{.6\linewidth}
\begin{align}
(GK)'' = G[-2i\Gamma_V(GK)' - \Gamma_T GK + K''].\\ \nonumber
\end{align}
\end{minipage}
\normalsize
Gaussian Wuppertal smearing is given by \small$K(x,y;\mathbf p) = \sum_{x',x'',...} \underbrace{K_0(x,x';\mathbf p) K_0(x',x'';\mathbf p) ... K(x^{'...'},y;\mathbf p)}_{N_W}$, \normalsize with
%\small
\begin{align}
K_0(x,y;\mathbf p) &= e^{-i\mathbf {p(x-y)}} \frac{1}{1+6\alpha} \Bigg(\delta_{x,y} + \alpha\sum_{j=1}^3 \Big[\ U_j(x) \delta_{x+\hat{\jmath},y}  + U_j^\dagger (x-\hat{\jmath}) \delta_{x-\hat{\jmath},y}\Big]\Bigg)\nonumber \\
&= \frac{1}{1+6\alpha} \Bigg( \delta_{x,y} + \sum_{j=1}^3 \alpha \Big[ e^{ip^j} U_j(x) \delta_{x+\hat{\jmath},y} + e^{-ip^j} U_j^\dagger(x-\hat{\jmath}) \delta_{x-\hat{\jmath},y}\Big]\Bigg).
\end{align}
\normalsize
The $m$th derivative of $K_0$ at zero momentum is equal to
\begin{equation}
K_0^{(m)}(x,y) \equiv \Big( \frac{\partial}{\partial p^j}\Big)^m K_0(x,y;\mathbf p) \Bigg |_{\mathbf p=0}= \frac{\alpha}{1+6\alpha} \Bigg[ i^m U_j(x) \delta_{x+\hat{\jmath},y} + (-i)^m U_j^\dagger (x-\hat{\jmath}) \delta_{x-\hat{\jmath},y}\Bigg].
\end{equation}
The first and second derivatives of smearing with $N_W$ iterations, $K = K_0^{N_W}$, can be computed iteratively using $(K_0^N)' = K_0'K_0^{N-1} + K_0(K_0^{N-1})', \; (K_0^N)'' = K_0'' K_0^{N-1} + 2 K_0'(K_0^{N-1})' + K_0(K_0^{N-1})''.$
%----------------------
\section{Ground-state contribution to correlation functions}
\label{Sd}
%---------------------
We will be tracing the correlators with polarization matrices that contain the projector $(1+\gamma^0)/2$, so that we can effectively write the overlap matrix elements as $\langle 0|N_\alpha(0)|N(p,s)\rangle=Z(\mathbf p)u_\alpha(p,s)$, \cite{m,n} .
 Here and in the following we use Minkowski-space gamma matrices. The Dirac and Pauli form factors, $F_1^{(N,q)}(Q^2)$ and $F_2^{(N,q)}(Q^2)$, parametrize matrix elements of the vector current %between proton states
\begin{equation}
\langle N(p',s') | \bar q \gamma^\mu q | N(p,s)\rangle = \bar u(p',s') \mathscr{F}[\gamma^\mu, \mathbf{p'},\mathbf{p}] u(p,s),
\end{equation}
with  the short-hand notation $\mathscr{F}[\gamma^\mu,\mathbf{p'},\mathbf{p}] = F_1^{(N,q)} \gamma^\mu + F_2^{(N,q)} \; \frac{i\sigma^{\mu\nu}(p'-p)_\nu}{2m}$, 
where $Q^2 = -(p'-p)^2$. Having $T =|x^0 - z^0|$ and $\tau=|y^0-z^0|$, the ground-state contributions to the two- and three-point functions (for $\mathbf{p'}=0$) are:
\begin{align}
&C_2 (\mathbf p, T)_{\alpha\beta} = \frac{Z^2(\mathbf p)}{2E(\mathbf p)} e^{-E(\mathbf p) T}(m+\slashed p)_{\alpha\beta}, \\
&C_3(\mathbf p, \tau, T)_{\alpha\beta} = \frac{Z(0)}{2m}\frac{Z(\mathbf p)}{2E(\mathbf p)}  e^{-m(T-\tau)} e^{-E(\mathbf p) \tau } 2m \big[\mathscr F(\gamma^\mu,0,\mathbf p) (m+\slashed p)\big]_{\alpha\beta}.
\end{align}
%with 
\normalsize
%---------------------
\section{Momentum derivatives of the ratio}
\label{MD}
%------------------------
Because we don't know how $Z(\mathbf p)$ depends on the momentum, we need to compute derivatives of the ratio of three-point and two-point functions. We set $\mathbf p'=0$ and $\mathbf p = k \mathbf e_j$, where $\mathbf e_j$ is the unit vector in $j$-direction. We compute the following ratio:
%\small
\begin{equation}
R_{\alpha\beta}(k,\tau,T) = R_N(k,\tau,T)_{\alpha\beta}\;R_A(k,\tau,T), \;\text{with}
\end{equation}
\begin{equation}
R_N(k,\tau,T)_{\alpha\beta} = \frac{C_3(k,\tau,T)_{\alpha\beta}}{ \sqrt{C_2(\mathbf p'=0,T) C_2(k,T)}},\; R_A(k,\tau,T) = \sqrt{\frac{C_2(k,T-\tau) C_2(\mathbf p'=0,\tau)}{C_2(\mathbf p'=0,T-\tau)C_2(k,\tau)}}.
\end{equation}
\normalsize
For computing the first and second momentum derivatives of this ratio we need:
\small
\begin{align}
R_N'(k)_{\alpha\beta} &= \frac{-C'_2(k)C_3(k)_{\alpha\beta} + 2 C_2(k)C'_3(k)_{\alpha\beta}}{2\sqrt{C_2(0) C_2(k)^3}}, \label{fd}\\
R_N''(k)_{\alpha\beta} &= \frac { (3[C'_2(k)]^2 - 2C_2(k) C''_2(k))C_3(k)_{ \alpha\beta} + 4C_2(k) (-C'_2(k)C'_3(k)_{\alpha\beta} + C_2(k)C''_3(k)_{\alpha\beta})}{4\sqrt{C_2(0)C_2(k)^5}}.\label{sd}
\end{align}
\normalsize
where, for more readability we suppress $\tau, T$ parameters, use the notation $C_2(k) = \text{Tr}\big(\frac{1+\gamma^0}{2}\; C_2(k)\big)$ and denote the derivatives with a prime e.g. $C'_2(k) \equiv \frac{dC_2(k)}{dk} $. 
We know that $C'_2(0) = 0$ in the infinite-statistics limit. Hence, we can eliminate this from the ratios.
Similarly, we can calculate $R'_A(k)$ and $R''_A(k)$ which can be used together with \ref{fd} and \ref{sd} to calculate the first and second derivatives of the ratio $R_{\alpha\beta}$. These derivatives are computed on the lattice directly at $k=0$ as discussed earlier in section \ref{der}.
The ground-state contributions are equal to: 
\small
\begin{align}
R_{\alpha\beta}(k) &= \frac{[\mathscr F(k) (m+E\gamma^0-k\gamma^j)]_{\alpha\beta}}{8\sqrt{2E(E+m)}}, \\
R'_{\alpha\beta}(k)  &= \frac{\big[ \mathscr F'(k)(m+E\gamma^0-k\gamma^j)\big]_{\alpha\beta} + \big[\mathscr F(k) (E'\gamma^0-\gamma^j)\big]_{\alpha\beta}}{8\sqrt{2E(E+m)}} - \frac{ \big[\mathscr F(k) (m+E\gamma^0 -k\gamma^j)\big]_{\alpha\beta} (2E+m)E' }{16\sqrt 2 [E(E+m)]^{3/2}},
\end{align}
\normalsize
and $R''_{\alpha\beta}(k)$ can be calculated in a similar way.
We use the continuum dispersion relation $E(k) = \sqrt{m^2+k^2}$, which implies $Q^2 = 2m\sqrt{m^2+k^2} -2m^2$, and find that at $k=0$, the second derivative is needed to obtain the slope of $F_1$:
\small
\begin{equation}
\frac{dF_1}{dk}\Big|_{k=0} = \frac{dQ^2}{dk}\Big|_{k=0}\frac{dF_1}{dQ^2}\Big|_{Q^2=0} = 0, \qquad
 \frac{d^2F_1}{dk^2}\Big|_{k=0} = 2 \frac{dF_1}{dQ^2}\Big|_{Q^2=0}.
\end{equation}
\normalsize
 Furthermore, we have at $k=0$, $E(0) =m,\; E'(0) = 0, \; E''(0) = 1/m$ and, 
\begin{equation}
\mathscr F(0) = F_1(0)\gamma^\mu, \quad \mathscr F'(0) = F_2(0)\frac{i\sigma^{\mu j}}{2m},  \quad   \mathscr F''(0) = 2\frac{d}{dQ^2} F_1(0) \gamma^\mu - F_2(0) \frac{i\sigma^{\mu0}}{2m^2}.
\end{equation}
%Finally, we evaluate the ratios and their derivatives at $k=0$,
Using $\Gamma_{pol} = (1+\gamma^3\gamma_5) \frac{1+\gamma^0}{2}$, we find nonzero results for the following values of the index $\mu$ (labeling the components of the vector current),
\small
\begin{align}
&\text{Tr}[R(\mu=0)\Gamma_{pol}] = F_1, \quad \text{Tr} [\partial_1 R(\mu=2)\Gamma_{pol}] = -\frac{i}{2m}(F_1 + F_2), \label{eq1}\\
&\text{Tr} [\partial_2 R(\mu=1)\Gamma_{pol}] =\frac{i}{2m} (F_1+F_2), \;\;
\text{Tr}[\partial_{1,2,3}^2 R(\mu=0) \Gamma_{pol}] = -\frac{1}{4m^2} (F_1+2F_2) - \frac{1}{3} F_1r_1^2, \label{eq2}
\end{align}
\normalsize
with $\partial_i = \frac{\partial}{\partial p^i}$ and  $r_1^2 = \frac{-6}{F_1}\quad \frac{dF_1}{dQ^2}\Big|_{Q^2=0}$. From equations \ref{eq1} and \ref{eq2} we find the following relations for the anomalous magnetic moment $\kappa$ and Dirac radius $r_1$:
%\small
\begin{align}
\kappa &= -2\,m\, \text{ Im}(\text{Tr}[R'(\mu=2)\;\Gamma_{pol}]) -  \text{Tr}[R(\mu=0)\;\Gamma_{pol}],  \label{k}\\
r_1^2 &= \frac{12\,m\, \text{Im}[R'(\mu=2)\;\Gamma_{pol}] + 3\, \text{Tr}[R(\mu=0)\;\Gamma_{pol}] -12\,m^2\, \text{Tr}[R''(\mu=0)\;\Gamma_{pol}] }{4\,m^2\, \text{Tr}[R(\mu=0)\;\Gamma_{pol}]}, \label{r2}
\end{align}
\normalsize
where we average over equivalent vector components and directions:
\small
\begin{align}
&\text{Tr}[R'(\mu=2)\Gamma_{pol}] = \frac{1}{2}(\text{Tr}[\partial_1R(\mu=2)\;\Gamma_{pol}] - \text{Tr}\nonumber [\partial_2R(\mu=1)\;\Gamma_{pol}]).\\
&\text{Tr}[R''(\mu=0)\Gamma_{pol}] = \frac{1}{3}(\text{Tr}[\partial_1^2R(\mu=0)\;\Gamma_{pol}] + \text{Tr}[\partial_2^2R(\mu=0)\;\Gamma_{pol}] +  \text{Tr}[\partial_3^2R(\mu=0)\;\Gamma_{pol}]).
\end{align} 
\normalsize
%-------------------
\section{Results and conclusions}
\label{LC}
\begin{figure}
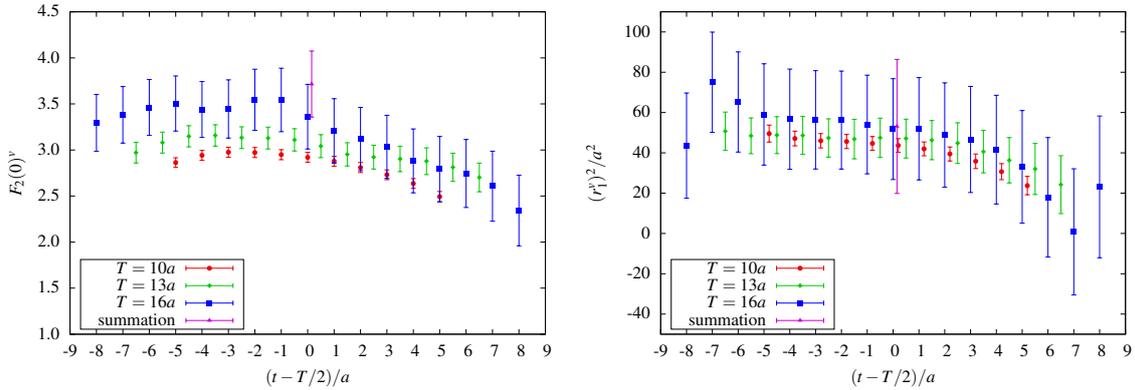
%[H]
  \resizebox{0.5\linewidth}{!}{\input{F2_2.tex}} \hfill
  \resizebox{0.5\linewidth}{!}{\input{r2_LS_2.tex}} 
  \caption{Anomalous isovector magnetic moment (left) and isovector Dirac radius (right). For both $\kappa$ and $[r_1^v]^2/a^2$, results from ratio method are shown using source-sink separations $T/a \in \{10, 13, 16\}$, as well as the summation method.}
  \label{fig1}
\end{figure}

\begin{figure}
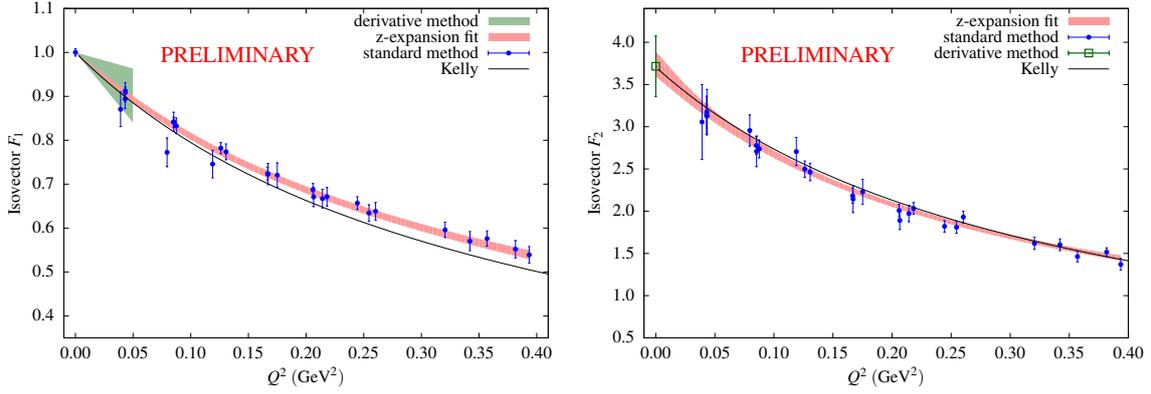
%[H]
  \resizebox{0.5\linewidth}{!}{\input{F1_summ2_2.tex}} \hfill
  \resizebox{0.5\linewidth}{!}{\input{F2_summ2_2.tex}} 
  \caption{Isovector Dirac (left) and Pauli (right) form factors. The blue points show results from
the standard method and the red bands show a z-expansion fit to those points. The green band (left) and
point (right) show the slope and value of the respective form factor at $Q^2=0$ , computed using
the momentum derivative method. The black curves result from a phenomenological fit to
experimental data by Kelly \cite{q}.}
\label{fig2}
\end{figure}

%---------------------
We perform lattice QCD calculations using a tree-level Symanzik-improved gauge action \cite{p,ss} and 2+1 flavors of tree-level improved Wilson-clover quarks, which couple to the gauge links via two levels of HEX smearing, at the physical pion mass $m_\pi=135$ MeV, lattice spacing $a=0.093$ fm, and a large volume $L_s^3 \times L_t= 64^4$ satisfying $m_\pi L = 4$. %For a detailed description of the action and smearing procedure we refer the reader to \cite{p}.
We are measuring the isovector combination $u-d$ of the three-point functions, where the disconnected contributions cancel out.
% For computing the three-point function we use sequential propagators through the sink.
 Furthermore, we perform measurements using three source-sink separations $T/a\in\{10, 13,16\}$ ranging from $0.9$ fm to $\sim 1.5$ fm, and we are using the summation method \cite{k,l} for removing contributions from excited states. We apply our analysis on $442$ gauge configurations, using all-mode-averaging  \cite{z} with $64$ approximate samples and one high-precision bias correction per configuration.
  %and in order to reduce the large computational cost, we made use of the all-mode-averaging (AMA) error reduction technique \cite{z} with $64$ sources (low-precision) and $1$ source (high-precision)  per gauge configuration for bias correction. 
The ``plateau plots'' in Figure \ref{fig1}, show $F_2^{v}(0)$ (left-hand side, using \ref{k}) and $[r_1^v]^2/a^2$ ( right-hand side, using \ref{r2}) from the ratio method, as well as the summation method.  
%$F_2^{v}(0)$, that we calculate using relation \ref{k} is shown in the left-hand side of figure \ref{fig1} using the ratio method, as well as the summation method.
%for source-sink separations $T/a \in \{10,13,16\}$ and also using the summation method,
 %on the right-hand side of figure \ref{fig1} one can see the resulted $[r_1^v]^2/a^2$ using relation \ref{r2}.
Figure \ref{fig2} shows a comparison between the results we get for the anomalous magnetic moment $\kappa=F_2^{v}(0)$ and the isovector Dirac radius $[r_1^v]^2$ using the momentum derivative approach and what we get using the traditional approach of measuring Pauli and Dirac form factors for each value of $Q^2$ (on the same ensemble) and then applying the z-expansion fit \cite{i, j}.  Preliminary results are given in Table \ref{tab}.\\
We confirm that our approach produces results consistent with those obtained using the traditional method. However, we found that this approach yields large statistical uncertainties, especially for the Dirac radius, which requires two momentum derivatives applied to a single quark line. Therefore, putting one momentum derivative on each of two different quark lines, as was done in \cite{o}, might be less noisy.

 %We calculated the momentum derivatives only with respect to the initial-state quark. Hence, further improvement can be achieved by including the momentum derivatives with respect to the final-state quarks as well.

\begin{table}%[H]
\centering
\begin{tabular}{|P{2.5cm}|P{3.5cm}|P{3.5cm}|}
\hline
\text{Quantity} & Traditional method & Derivative method\\
\hline
$\kappa^v$ & 3.74(14) & 3.71(35)\\
\hline
$[r_1^v]^2$ $[\text{fm}]^2$ &0.55(9) & 0.45(28)\\
\hline
\end{tabular}
\caption{A comparison between the derivative method and the traditional one}
\label{tab}
\end{table}

%----------------
%---------------------------
%\section{Conclusion}
%The derivative method is a model independent approach which enables us to measure the Dirac radius directly at $Q^2=0$ and it delivers %results consistent with the results we get using the traditional method but with $2-3$ times larger statistical uncertainties, for that we plan %to extend our study by calculating the momentum derivatives with respect to the sink momentum too.
%---------------------------
\small
\acknowledgments
We thank the Budapest-Marseille-Wuppertal collaboration for making their configurations  \cite{p,ss} available to us.  This research used resources at Forschungszentrum J\"ulich and on CRAY XC40 (Hazel Hen) at HLRS. SM is supported by NSF grant PHY-1520996,  SM and SS are supported by  RBRC, JN is supported by the Office of Nuclear Physics of the U.S. Department of Energy (DOE) under Contract DE-SC0011090, ME is supported by DOE grant DE-FG02-96ER40965 and AP is supported in part by DOE under grant DE-FC02-06ER41444.\\
Calculations for this project were done using the Qlua software suite \cite{s}.

\end{document}